\documentclass[%
 aps,
 pre,%
 amsmath,amssymb,
 reprint,%
superscriptaddress,
]{revtex4-1}

\usepackage{graphicx}% Include figure files
\usepackage{dcolumn}% Align table columns on decimal point
\usepackage{bm}% bold math
\usepackage{color}

\begin{document}
\newcommand{\micron}{$\mu$m}

\title{Theoretical and experimental investigation of the equation of state of boron plasmas}
\author{Shuai Zhang}
\email{zhang49@llnl.gov}
\affiliation{Lawrence Livermore National Laboratory, Livermore, California 94550, USA}
\author{Burkhard Militzer}
\email{militzer@berkeley.edu}
\affiliation{Department of Earth and Planetary Science, University of California, Berkeley, California 94720, USA}
\affiliation{Department of Astronomy, University of California, Berkeley, California 94720, USA}
%\affiliation{Lawrence Livermore National Laboratory, Livermore, California 94550, USA}
\author{Michelle C. Gregor}
\email{gregor3@llnl.gov}
\author{Kyle Caspersen}
\author{Lin H. Yang}
\author{Tadashi Ogitsu}
\author{Damian Swift}
\author{Amy Lazicki}
\author{D. Erskine}
\author{Richard A. London}
%\author{D. E. Fratanduono}
\author{P. M. Celliers}
%\author{F. Coppari}
%\author{J. H. Eggert}
\author{Joseph Nilsen}
\author{Philip A. Sterne}
\author{Heather D. Whitley}
\email{whitley3@llnl.gov}
\affiliation{Lawrence Livermore National Laboratory, Livermore, California 94550, USA}

\date{\today}

\begin{abstract}
{
We report a theoretical equation of state (EOS) table for boron across a wide range of temperatures
(5.1$\times$10$^4$--5.2$\times$10$^8$ K) and densities (0.25--49 g/cm$^3$),
and experimental shock Hugoniot data at unprecedented high pressures (5608$\pm$118 GPa).
The calculations are performed with full, first-principles methods combining path integral Monte Carlo (PIMC) at high temperatures and density functional theory molecular dynamics (DFT-MD) 
methods at lower temperatures.
PIMC and DFT-MD cross-validate each other by providing coherent EOS 
(difference $<$1.5 Hartree/boron in energy and $<$5\% in pressure) at 5.1$\times$10$^5$ K.
The Hugoniot measurement is conducted at the National Ignition Facility using a planar shock platform.
The pressure-density relation found in our shock experiment is on top of the 
shock Hugoniot profile predicted with our first-principles EOS and a semi-empirical EOS table (LEOS 50).
We investigate the self diffusivity and 
the effect of thermal and pressure-driven ionization on the EOS and shock compression behavior in high pressure and temperature conditions
We study the performance sensitivity of a polar direct-drive exploding pusher platform to pressure variations based on comparison 
of the first-principles calculations with LEOS 50 via 1D hydrodynamic simulations.  
The results are valuable for future theoretical and experimental studies and engineering design in 
high energy density research. (LLNL-JRNL-748227)}
\end{abstract}

%\pacs{62.50.−p, 31.15.A−, 61.20.Ja, 64.30.−t}


\maketitle

\section{Introduction}
Recent experiments at the National Ignition Facility (NIF) have demonstrated the utility of large diameter polar direct-drive exploding pushers (PDXP) as a low areal density platform for nucleosynthesis experiments,\cite{Maria_2018} 
neutron source development, neutron and x-ray diagnostic calibration, and potentially as a candidate platform for heat transport studies.\cite{Ellison_2018}  Improving the platform for each 
of these respective uses requires consideration of various model uncertainties.  Achieving a lower shell areal density during burn or obtaining additional data to help constrain estimates of this quantity in the nucleosynthesis experiments 
would simplify analysis of the charged particle data collected, while improving implosion symmetry is a necessary requirement if the platform is to be used to study heat transport.  Variations in the ablators used in these experiments is 
one possible avenue that is currently under investigation.  The use of glow-discharge polymer (GDP) as an ablator improves performance over smaller glass capsules,\cite{Maria_2018} but its low tensile strength 
requires designs with shell thickness of about 15-20 $\mu$m in order to support gas fill pressures of around 8 bar.  Higher tensile strength materials offer the option of producing thinner shells to support similar fill 
pressures, and 
reactions of ablator materials with neutrons and protons could potentially be used to obtain additional data to help quantify shell areal density at burn time.  Some candidate materials with higher tensile strength  include beryllium, boron, boron carbide, 
 boron nitride, and high density carbon.  For the purpose of conducting heat flow measurements, beryllium was ruled out as a candidate 
material due to the inclusion of argon within the capsule during the fabrication process.\cite{Ellison_2018}
Boron and nitrogen, which both undergo reactions with neutrons and protons, offer the potential 
for using additional nuclear reactions to better constrain the shell areal density during nuclear burn time, which could improve our overall understanding of the effects of the shell on the measured charged particles in the nucleosynthesis experiments.  Boron is also interesting as an ablator material since its reactions with $\gamma$-rays could be used to constrain 
ablator mix at burn time.\cite{LA-UR-15-20627}  

Radiation hydrodynamic simulations are the workhorse method for design and analysis of the inertial confinement fusion (ICF) and high energy density experiments.  It has been demonstrated in many previous studies that the equation of state (EOS) of capsule ablator 
materials is an important component in indirect drive ICF performance,\cite{HAMMEL2010171,PhysRevLett.111.065003,PhysRevLett.112.185003,Hu_2015,PhysRevE.92.043104} and EOS may also affect the implosion dynamics in the polar direct-drive platform, impacting not only capsule yield, but also the shell areal density during burn 
and the electron-ion temperature separation in the gas.  Thus, exploration of these materials as candidates for future PDXP-based experiments requires reasonable EOS models for use in radiation hydrodynamic simulations.  In this paper, we examine the EOS of boron via both {\it ab initio} simulations and experimental measurements.  We also examine its performance as an ablator in 1D simulations of the PDXP platform, focusing on how variations in the EOS impact the computed yield and plasma conditions at burn time.

EOS models that are widely used in hydrodynamic simulation codes, 
such as the quotdian EOS (QEOS)~\cite{leos1qeos,leos2}, 
provide pressures and energies as smooth functions of
temperature and density based on semi-empirical methods,
such as the Thomas-Fermi (TF) theory.
The TF theory treats the plasma as a collection of nuclei that follow Boltzmann statistics 
and electrons that form continuous fluids and obey Fermi-Dirac statistics.
This offers a good means to describe weakly-coupled plasmas and 
materials at very high densities,
but is insufficient in describing many condensed matter solids and liquids,
where bonding effects are significant.  
Additionally, at low-to-intermediate temperatures where atoms
undergo partial ionization, the TF theory does not accurately capture 
the effects of shell ionization, which impacts the electronic contribution to the EOS of the material.

There has been continuous research in the development of improved methods for computing thermodynamic properties of materials, 
which has resulted in a variety of methods that can be applied to study EOS across a wide range of densities and temperatures.  
Methods appropriate to the study of plasmas include density functional theory (DFT)-based methods such as INFERNO~\cite{Inferno}, Purgatorio~\cite{Purgatorio2006,Sterne2007}, orbital-free (OF) quantum molecular dynamics (MD)~\cite{ofmd,Danel2012},
and extended-DFT~\cite{ZhangExtendedDFT2016},
activity-expansion method (ACTEX)~\cite{Rogers86,Rogers94,Rogers96}, 
and many-body path integral Monte Carlo (PIMC)~\cite{Pollock1984,pollock1988,Ceperley1995,Ce96,Militzer2015} methods.
Standard Kohn-Sham DFT-MD has been widely applied for EOS studies 
of condensed matter as well as warm and hot, dense plasmas.
It accounts for both the electronic shells and bonding effects, and is thus superior 
 to average-atom methods in situations where these types of strong many-body correlations are impactful to the EOS.
However, the DFT-MD approach
becomes computationally intractable at high temperatures
because considerable numbers of partially occupied orbitals need to be
considered.

As a powerful tool initially developed for hydrogen~\cite{PhysRevLett.73.2145},
PIMC has been successfully utilized to study plasmas from weak coupling 
to strongly coupled regimes with high accuracy.
Recent developments by Militzer et al.~\cite{Driver2012,Militzer2015}
provide useful recipies for studying higher-Z plasmas.
In the past seven years, they have implemented the PIMC methods under the
fixed-node approximation and obtained the EOS for a series of elements
(C, N, O, Ne, Na, Si)~\cite{Driver2012,Militzer2015,Driver2016Nitrogen,Driver2015Oxygen,Driver2015Neon,Zhang2016b,Zhang2017}
and compounds (H$_2$O, LiF, hydrocarbons)~\cite{Driver2012,Driver2017LiF,Zhang2017b,Zhang2018}
over a wide range of temperature, pressure conditions.
The goal of the theoretical part of this paper is to apply these methods to calculate the EOS of boron,
and explore the effect on PDXP simulations in comparison with an older EOS model (LEOS 50) 
through hydrodynamic simulations.

Located in between metals and insulators in the periodic table,
the structure and properties of boron have attracted wide interest
in high pressure physics.
A number of studies have examined 
the stability relations of the $\alpha$ and the $\beta$ phases~\cite{Masago2006,Shang2007,Ogitsu2013}.
A phase diagram was proposed for crystalline boron based on 
DFT simulations~\cite{Oganov2009}, showing five different phases 
at pressures up to 300 GPa, 
part of which having been confirmed in static compression experiments using diamond anvil cells.
A considerable amount of study has been performed on boron at low densities,
including DFT-MD simulations and X-ray radiography measurements
on the structure, electronic, and thermodynamic properties of liquid
boron~\cite{Vast1995,Clerouin2008,Price2009}, general chemical models
for the the composition and transport properties of
weakly-coupled boron plasmas~\cite{Apfelbaum2013}, 
isochoric EOS and resistivity of warm boron by combining 
closed vessel experiments, DFT-MD, average-atom methods, 
and a chemical model (COMPTRA)~\cite{Clerouin2012,Clerouin2010,JOHNSON2006327,Blancard2004,Faussurier2010,comptra2005}.
In comparison to the vast progress in the low-temperature, high-pressure  
and the high-temperature, low-pressure regions of the boron phase diagram,
studies at simultaneously high pressures and temperatures are rare.
Until the year 2013, the only shock Hugoniot data available were at pressures 
below 112 GPa~\cite{MarshLASL1980}.
Recently, Le Pape et al.~\cite{lepape2013} used X-ray radiography to study 
the structure of shocked boron. They reported two experimental Hugoniot
measurements and ion-ion structure factors that are consistent with 
DFT-MD simulations.
This extended the shock Hugoniot measurements of boron to the 
highest pressure of 400 GPa.

Hydrodynamic simulations of PDXP experiments require  the EOS 
of the ablator materials along and off
the Hugoniot curve at higher temperatures and pressures.  The
LEOS~\cite{leos1qeos,leos2} and SESAME~\cite{sesame} EOS databases may be
used, but it is unclear how their deviation from the true
values affect the reliability of results in PDXP simulations,
such as the neutron yield.
In this work, we perform calculations of the boron EOS over a wide range of 
temperatures and pressures.
We extend PIMC simulations of dense boron plasmas from the ``hot''
down to the ``warm'' region, where significant partial ionization
of the K shell persists and standard DFT-MD simulations with frozen 1s core
pseudopotentials are not trustworthy. At relatively low 
temperatures, the system behaves like the usual condensed matter
fluid, which can be reasonably well described within the
DFT-MD framework. By pushing PIMC to low temperatures
and DFT-MD to high temperatures, we get a
coherent, first-principles EOS table for boron.
We compare this table and the predicted shock compression
profiles with LEOS and SESAME EOS tables for boron,
and perform hydrodynamic simulations to compare the
effect of the different tables on the ICF performance.

The paper is organized as follows: Section~\ref{method} introduces the details of
our simulation methods and experiment.  Sec.~\ref{results} presents our EOS results, 
the calculated and measured shock Hugoniot data, 
and comparisons with other theories and models. 
Sec.~\ref{discuss} discusses 
the atomic and electronic properties of boron plasmas,
the ionization process,
and PDXP performance sensitive to the EOS;
finally we conclude in Sec.~\ref{conclusion}.

\section{Theory and experiment}\label{method}
\subsection{First-principles simulation methods}

Following the pioneering work applying PIMC to the simulations of
real materials (hydrogen)~\cite{PhysRevLett.73.2145} and recent 
development for pure carbon~\cite{Driver2012},
hydrocarbons~\cite{Zhang2017b,Zhang2018}, and lithium in LiF~\cite{Driver2017LiF}, 
our PIMC simulations~\cite{militzerphd}
 utilize the fixed-node approximation~\cite{Ceperley1991} 
and treat both electrons and the nuclei as quantum paths 
that are cyclic in imaginary time [0,$\beta$=$1/k_\text{B}T$], 
where $k_\text{B}$ is the Boltzmann constant.
We use free-particle nodes to constrain the path to positive regions of the
trial density matrix, which has been shown to work well for calculations of 
hydrogen~\cite{PhysRevLett.73.2145,PhysRevLett.76.1240,PhysRevE.63.066404,PhysRevLett.87.275502,PhysRevLett.104.235003,PhysRevLett.85.1890,Hu2011,CTPP:CTPP2150390137,Militzer20062136},  helium~\cite{PhysRevLett.97.175501,Mi05}, and other first-row elements~\cite{Driver2017LiF,Driver2012,Driver2015Oxygen,Driver2016Nitrogen,Driver2015Neon}.
The Coulomb interactions are described via pair density
matrices~\cite{Na95,pdm}, which are evaluated at an
imaginary time interval of [512 Hartree (Ha)]$^{-1}$. The
nodal restriction is enforced in much smaller steps of [8192 Ha]$^{-1}$.

For our DFT-MD simulations, we choose the hardest available projected augmented wave (PAW)
pseudopotentials~\cite{Blochl1994} for boron with core radii
of 1.1 Bohr and frozen 1s$^2$ electron, as provided in
the Vienna \textit{Ab initio} Simulation Package
({\footnotesize VASP})~\cite{kresse96b}.
We use the
Perdew-Burke-Ernzerhof (PBE)~\cite{Perdew96} functional 
to describe the electronic exchange-correlation interactions, 
which has been shown to be superior to the local density approximation
in studies of boron at low temperature~\cite{Widom2008}.
We choose a large cutoff energy of 2000 eV for the plane-wave basis, and we use the
$\Gamma$ point to sample the Brillouin zone.  The simulations are carried out in the 
 $NVT$ ensemble with a  temperature-dependent time step of 0.05-0.55 fs, chosen to 
 ensure reasonable conservation of energy.   
The temperature is regulated by a Nos\'{e} thermostat~\cite{Nose1984}.
Each MD trajectory typically consists of 5000 steps to ensure that the system has reached 
equilibrium 
and to establish convergence of the energies and pressures.
DFT-MD energies from {\footnotesize VASP} reported in this study are shifted by 
%-24.595631065~Ha/B, 
-24.596~Ha/B, 
the all-electron PBE energy of a single boron
atom determined with {\footnotesize OPIUM}~\cite{opium},
in order to establish a consistent comparison with the all-electron PIMC energies.

Our PIMC calculations are performed at temperatures from 
 5.05$\times$10$^5$ K to 5.17$\times$10$^8$ K 
and densities ranging from 0.1- to 20-times the
ambient density $\rho_0$ ($\sim$2.46 g/cm$^3$ 
based on that of the $\alpha$ phase~\cite{alphaB}).
We conduct DFT-MD simulations at temperatures between
5.05$\times$10$^4$~K and 10$^6$~K,
in order to check the PIMC calculations at the lowest temperatures. 
Due to limitations in applying the plane-wave basis for orbital expansion
at low densities, and limitations in the applicability of the pseudopotentials that freeze
1s$^2$ electrons in the core at high densities, we consider a smaller 
number of densities ($\rho_0$--10$\rho_0$) in DFT-MD. 
These conditions are relevant to the dynamic shock compression experiments we have conducted at the NIF,
and span the range in which Kohn-Sham DFT-MD simulations are feasible. 
All PIMC calculations use 30-atom cubic cells, while in DFT-MD 
we consider both 30-atom cells and larger cells with
108 and 120 boron atoms to
minimize the finite-size errors.

\begin{figure}
\centering\includegraphics[width=0.5\textwidth]{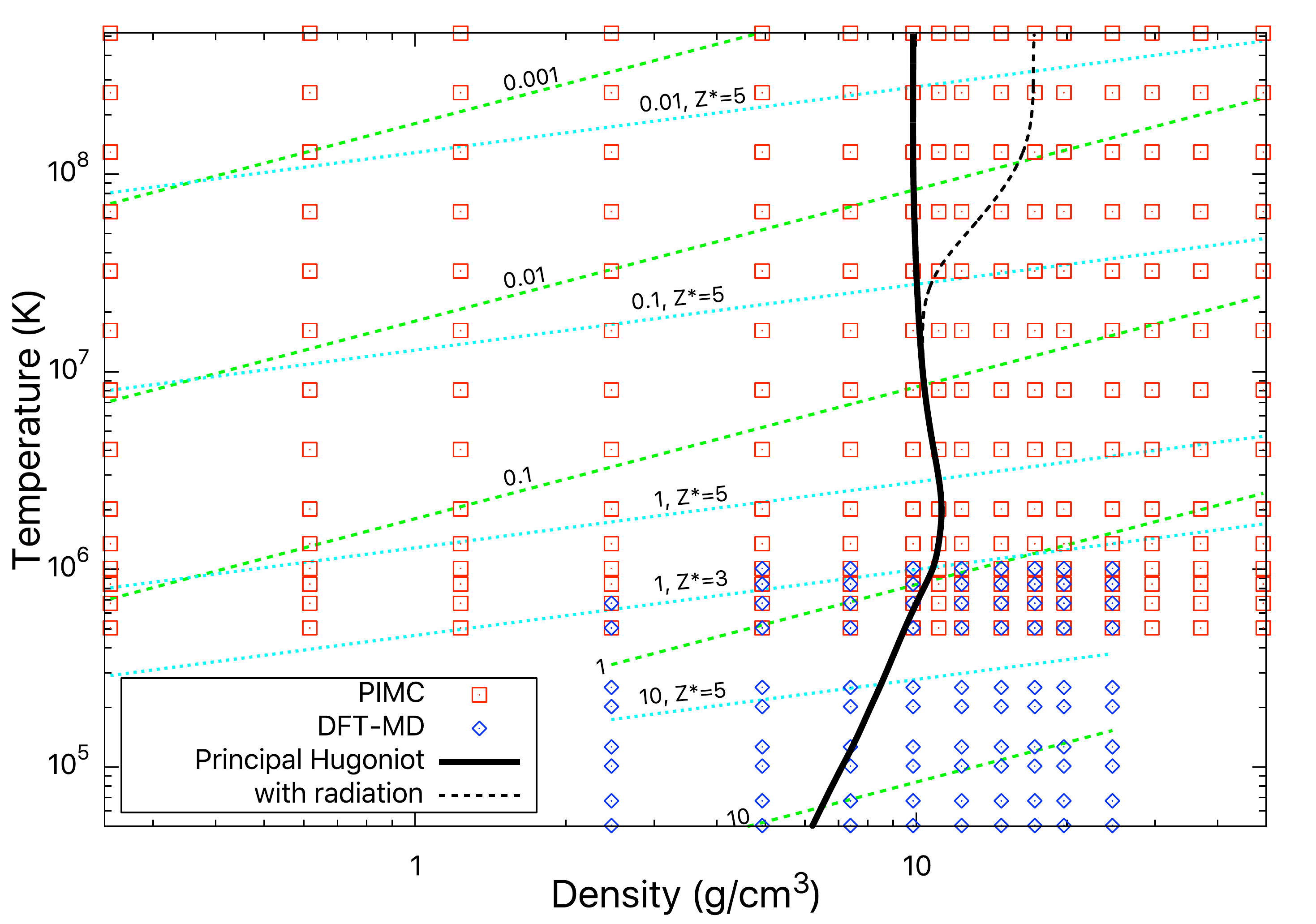}
\caption{\label{Trhogrid} Temperature-density conditions in our PIMC (red squares) and DFT-MD (blue diamonds) calculations are shown. The black curves depict the computed principal Hugoniot with (dashed) and without (solid) radiation correction~\cite{photoncorrection} to the EOS.
The dashed lines in green represent the conditions with different values of the degeneracy parameter, $\Psi$, and the dotted lines in cyan denote the effective ionic coupling parameter, $\Gamma$.
The Hugoniot curve is constructed by choosing the initial density to be the same as $\rho_0$ ($\sim$2.46~g/cm$^3$).}
\end{figure}

The temperature-density conditions included in this study are show in Fig.~\ref{Trhogrid}, along with contour lines corresponding to the ionic coupling parameter, $\Gamma=(Z^*e)^2/(ak_\text{B}T)$, and the 
electron degeneracy parameter, $\Psi=T_\text{Fermi}/T$, where $T_\text{Fermi}$ is the Fermi temperature of free electrons, $Z^*$ is the effective ion charge, $k_\text{B}$ is the Boltzmann constant, 
$a=(3/4\pi n)^{1/3}$ is the average ionic distance, and $n$ is the ion number density.
Our PIMC and DFT-MD calculations span a wide range of conditions for the boron plasma,
including weakly coupled ($\Gamma<1$) plasmas, as well as collisional, 
strongly coupled ($\Gamma>1$) and degenerate ($\Psi>1$) plasmas.  
We utilize the simulation data to predict the
 principal shock Hugoniot profile 
over a range of pressures spanning 10 to 10$^{5}$ megabar (Mbar),
as described in Section~\ref{secshock}.

\subsection{Shock Hugoniot experiment}
\begin{figure}
	\includegraphics[ width=3.375in, keepaspectratio]{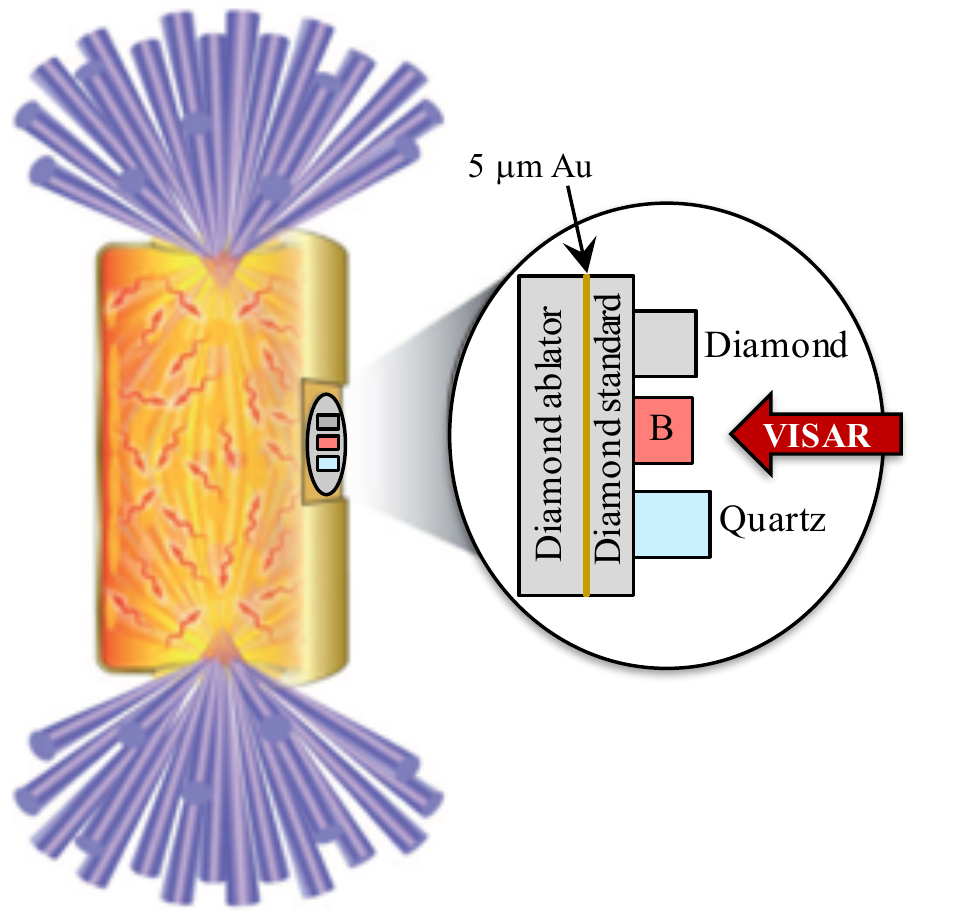}
	\caption{\label{fig:target} Target design for the impedance-matching experiment at the NIF.}
\end{figure}

An experiment to measure boron's Hugoniot near 50~Mbar was done at the NIF~\cite{NIF} at Lawrence Livermore National Laboratory (shot number N170801), using the impedance-matching technique. As shown in Fig.~\ref{fig:target}, the target physics package was affixed to the side of a gold hohlraum and comprised a 200-\micron{}-thick diamond ablator, 5-\micron{}-thick gold preheat shield, and a 100-\micron{}-thick diamond impedance-matching standard backing individual diamond, boron, and quartz samples. The optical-grade chemical vapor deposition diamond was polycrystalline with a density of 3.515 g/cm$^3$. The z-cut $\alpha$-quartz and the boron had densities of 2.65 g/cm$^3$ and 2.31 g/cm$^3$, respectively. 176 laser beams in a 5-ns pulse with a total energy of 827~kJ produced an x-ray bath in the hohlraum with a peak radiation temperature of 250 eV as measured by Dante~\cite{dante2004}. The x rays launched a strong, planar and nearly steady shock wave, varying $~\pm$3\% from its average velocity in the boron, that drove the samples to high pressures and temperatures.

The boron Hugoniot measurement was determined by impedance matching using the inferred shock velocities in the boron sample and diamond standard. Average shock velocities were determined from shock transit times, measured using a line-imaging velocity interferometer system for any reflector (VISAR)~\cite{Celliers_2004_VISAR}, and the initial sample thicknesses, measured using a dual confocal microscope. The average velocities were further corrected for shock unsteadiness witnessed \textit{in situ} in the transparent quartz sample \cite{Fratanduono_Nonsteady, Gregor_diamond, Fratanduono_B4C}. The Hugoniot and release for the diamond standard were calculated using a tabular equation of state (LEOS 9061) created from a multiphase model based on DFT-MD and PIMC calculations~\cite{Benedict_2014}. The experimental Hugoniot data are given in Table~\ref{table:expdata}.

\begin{table*}% add [H] placement to break table across pages
	\caption{\label{table:expdata} Boron Hugoniot data from impedance matching (IM) with a diamond standard. Shock velocities ($U_{\text{s}}$) at the IM interface were measured \textit{in situ} using VISAR for quartz (Q) and inferred using the nonsteady waves correction for boron (B) and diamond (C). $U_{\text{s}}^{\text{C}}$ and $U_{\text{s}}^{\text{B}}$ were used in the IM analysis to determine the particle velocity ($u_{\text{p}}$), pressure ($P$), and density ($\rho$) on the boron Hugoniot.}
		\begin{ruledtabular} %ruledtabular
			\begin{tabular}{cccccc}
				$U_{\text{s}}^{\text{Q}}$ & $U_{\text{s}}^{\text{C}}$  &$U_{\text{s}}^{\text{B}}$ &$u_{\text{p}}^{\text{B}}$& $P^{\text{B}}$ & $\rho^{\text{B}}$\\
				(km/s) & (km/s) & (km/s) &  (km/s) & (GPa) & (g/cm$^3$) \\
				\hline
				55.18 $\pm$ 0.25 & 55.25 $\pm$ 0.74 & 58.71 $\pm$ 0.66 & 41.35 $\pm$ 0.82 & 5608 $\pm$ 118 & 7.811 $\pm$ 0.465  \\
			\end{tabular}
		\end{ruledtabular}
\end{table*}

\section{Results}\label{results}

\subsection{Equation of state}\label{seceos}

 \begin{figure*}
\centering\includegraphics[width=0.49\textwidth]{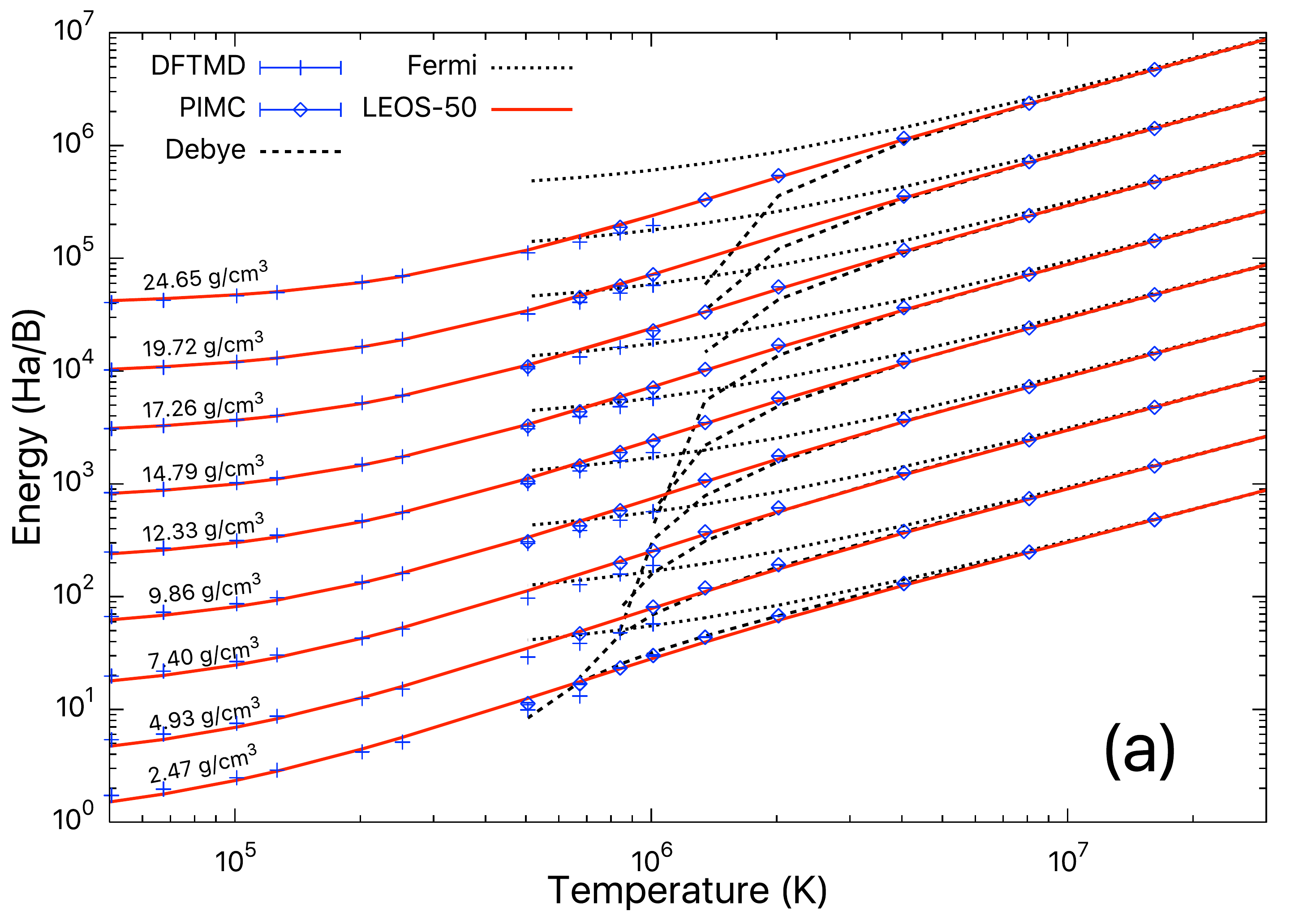}
\centering\includegraphics[width=0.49\textwidth]{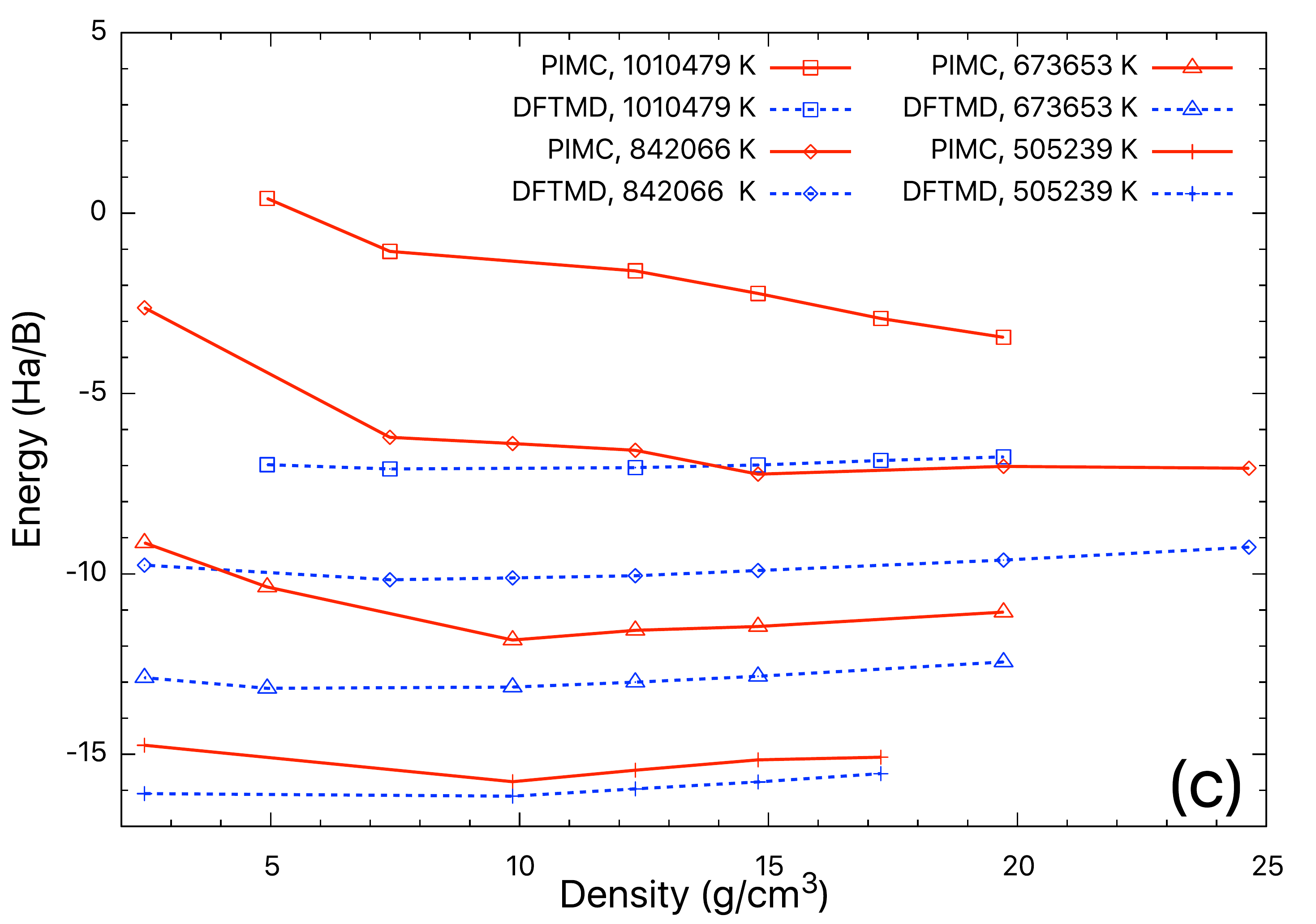}
\centering\includegraphics[width=0.49\textwidth]{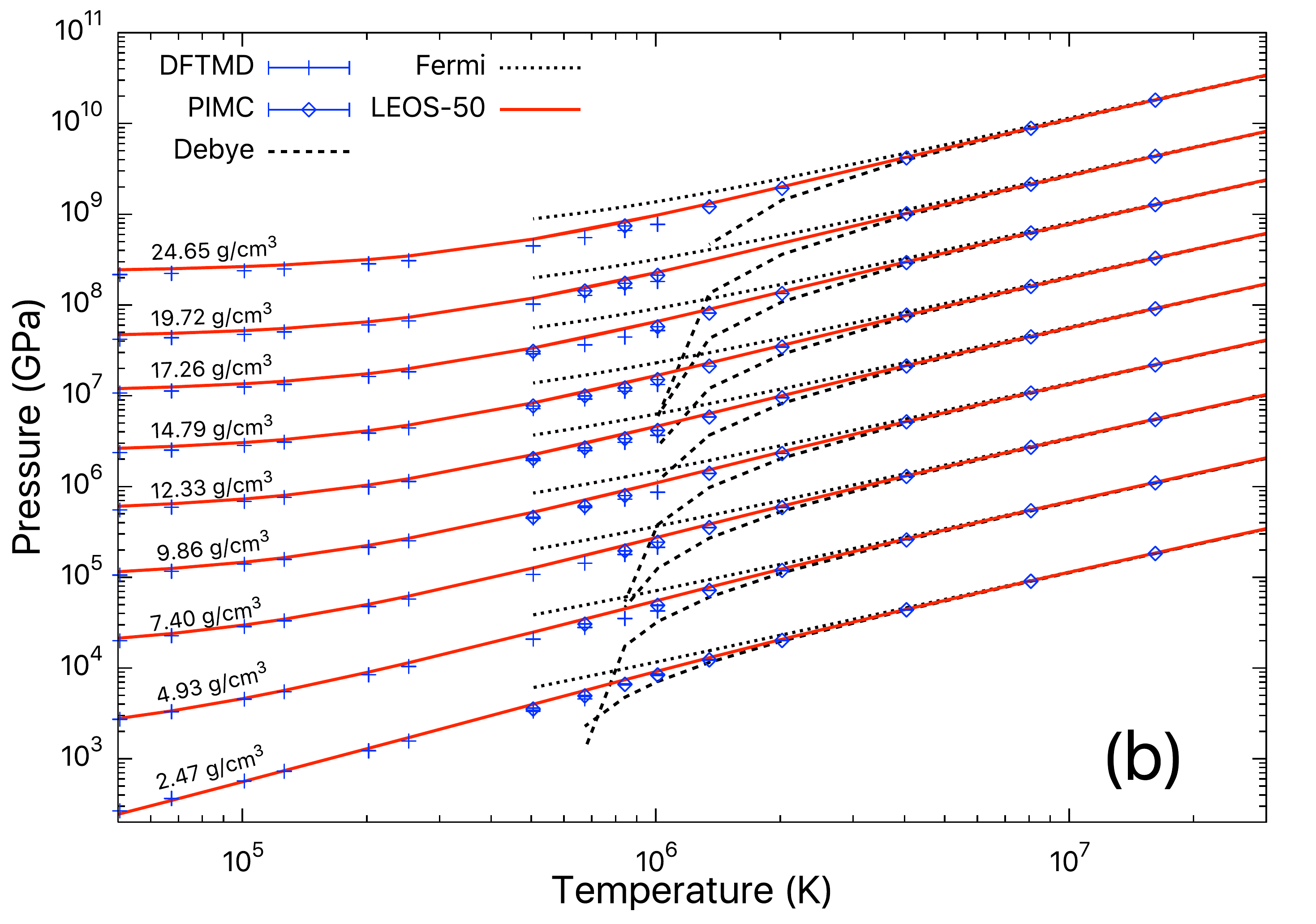}
\centering\includegraphics[width=0.49\textwidth]{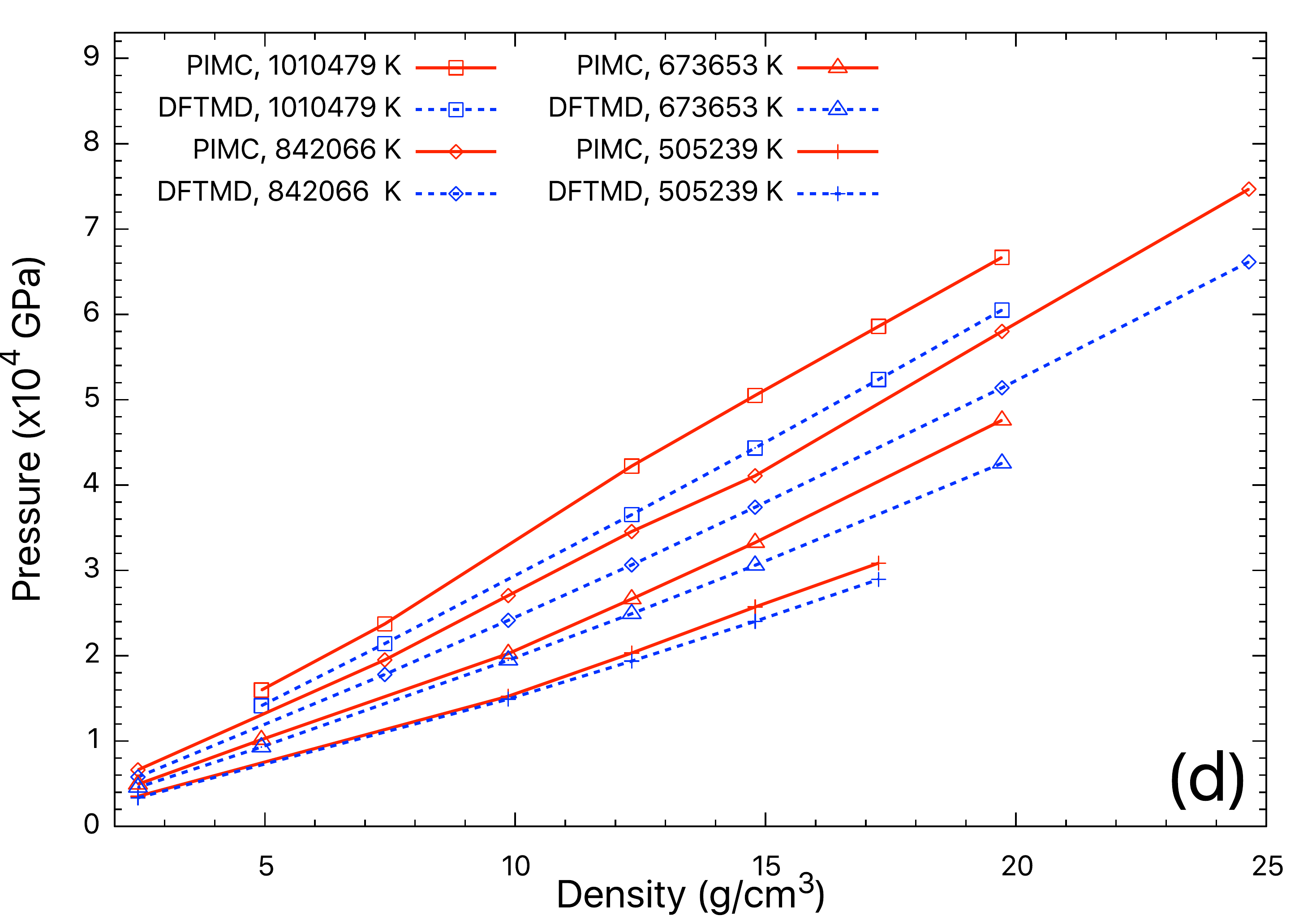}
\caption{\label{beosept} (a) Energy- and (b) pressure-temperature EOS plots along isochores 
for boron from our PIMC and DFT-MD simulations.  For comparison, the ideal Fermi-gas theory, Debye-H\"{u}ckel model, and LEOS 50 are also shown. In (a), the LEOS 50 data have been aligned with DFT by setting their energies to be equal at 2.46 g/cm$^3$ and 0 K. 
Subplots (c) and (d) are the comparison in internal energy and pressure between PIMC and DFT-MD along four isotherms as functions of density.
In subplots (a) and (b), results at different isochores have been shifted apart for clarity.}
\end{figure*}

The first-principles EOS computed with PIMC and DFT-MD calculations
are shown in Figs.~\ref{beosept}a and b. 
The internal energies
and pressures we computed using PIMC are consistent with those predicted by the
ideal Fermi gas theory and the Debye-H\"{u}ckel model in the high temperature limit ($>$1.6$\times$10$^7$~K) where these models are valid.
At lower temperatures, ideal Fermi gas theory and Debye-H\"{u}ckel model
predictions become increasingly higher and lower, respectively, than
our PIMC values for both internal energy and pressure.
This is easily understood due to the increased contribution from electron-electron and electron-ion 
correlations at lower temperature which render the high-temperature theories inadequate.
The PIMC energies and pressures show the same trend 
as those from our DFT-MD simulations along all the nine isochores 
between $\rho_0$--10$\rho_0$.

The explicit inclusion of electronic shell structures leads to
significant differences in the EOS of boron relative to
the TF model, in particular at $T\le2\times10^6$ K.
In comparison with our first-principles data, 
the LEOS 50 pressures differ by a variation -16.4\% to 7.1\%,
and the internal energy differences are between
-2.0--8.2 Ha/atom, at $T\le2.0\times10^6$ K. 
These differences lead to significantly different peak compression
in the shock Hugoniot curves, as will be discussed in Sec.~\ref{secshock}.
At high temperatures ($T>2\times10^6$ K), the relative differences in 
energies and pressures are small (between -3.1\% and 0.5\% in pressure, 
and between -1\% and 6\% in internal energy).

With decreasing temperature from 10$^6$ to 5.05$\times$10$^5$~K,
we find improved agreement between PIMC and DFT-MD results in both 
internal energy and pressure (Fig.~\ref{beosept}c,d). 
We define a critical temperature of 5.05$\times$10$^5$~K corresponding 
to the temperature above which significant ionization of the boron 
1s$^2$ core state is expected to 
render the pseudopotential calculation inaccurate.
This critical temperature is lower than what we found recently for
carbon in CH (10$^6$--2$\times$10$^6$~K). This is due to the shallower
1s level in boron than in carbon.
At the critical temperature, 
we find good consistency 
between PIMC and DFT-MD, with differences 
less than 1.5 Ha/B in energy and less than 5\% in pressure.

The larger underestimation in energy and 
pressure by DFT-MD at higher densities and temperatures can be attributed 
to the failure of the pseudopotential approximation at these conditions. 
The significant compression at densities higher than 5$\rho_0$ leads to the overlap of
the nearby frozen cores, which makes the use of the pseudopotential inaccurate at these conditions.  
In previous studies, 
other authors have overcome the failure of the pseudopotentials by constructing all-electron pseudopotentials that maintain accuracy up to higher temperatures and 
densities.\cite{Danel2012,Mazevet2007}  We note that the DFT-MD calculations shown here up to the critical temperature are in good agreement with the all-electron results, and the PIMC calculations 
agree with the all-electron calculations at the higher temperatures.

\subsection{Shock compression}\label{secshock}

During planar shock compression, the locus of the final (shocked) state
($E,P,V$) is related to the initial (pre-shocked) state ($E_0,P_0,V_0$) via 
the Rankine-Hugoniot equation~\cite{Meyers1994book}
\begin{equation}\label{eqhug}
(E-E_0) + \frac{1}{2} (P+P_0)(V-V_0) = 0,
\end{equation}
where $E$, $P$, and $V$ denotes internal energy, pressure, and volume, respectively.
Equation~\ref{eqhug} allows for determining the $P$-$V$-$T$ Hugoniot conditions 
with the EOS data in Sec.~\ref{seceos}. 

\begin{figure}
\centering\includegraphics[width=0.5\textwidth]{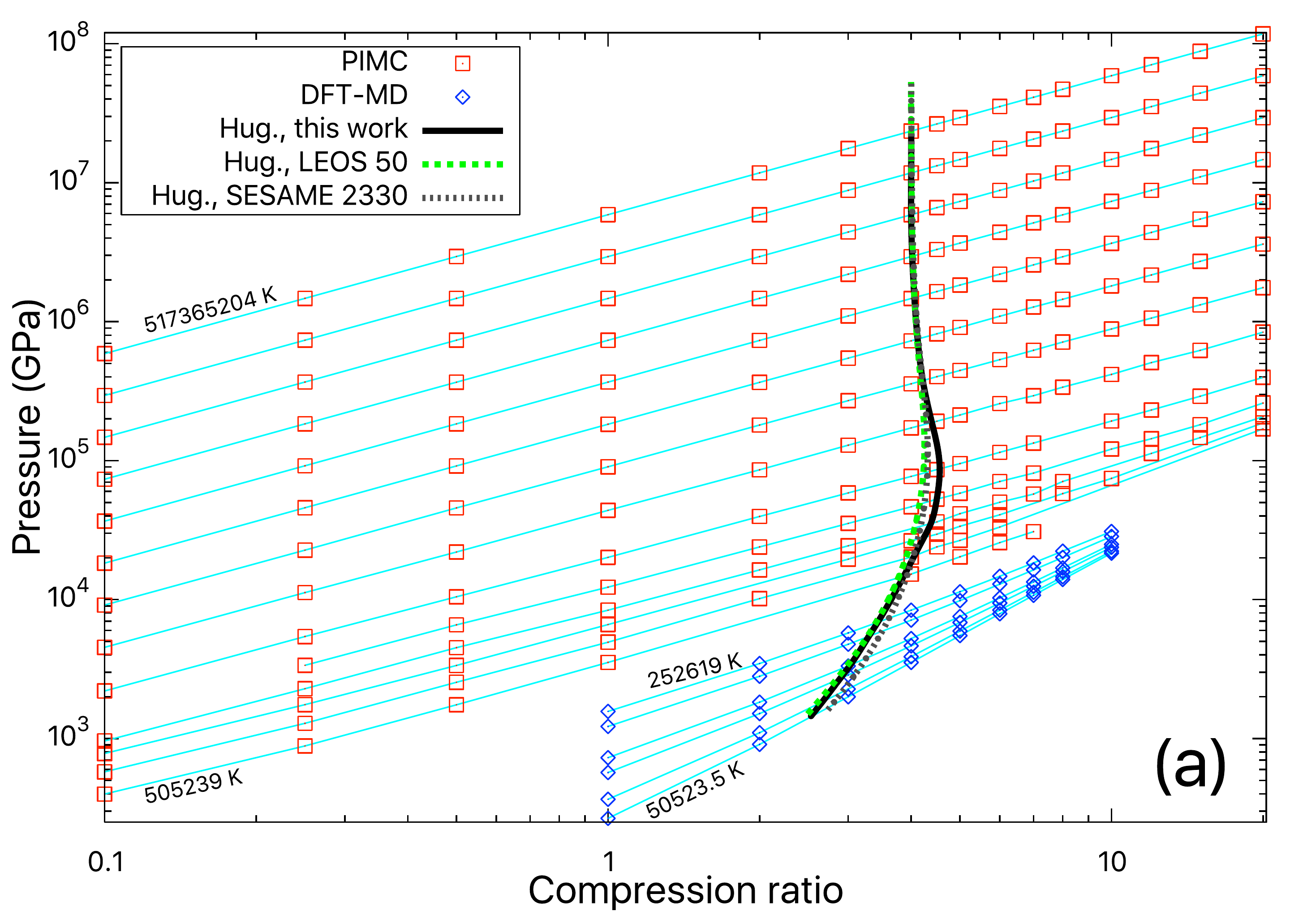}
\centering\includegraphics[width=0.5\textwidth]{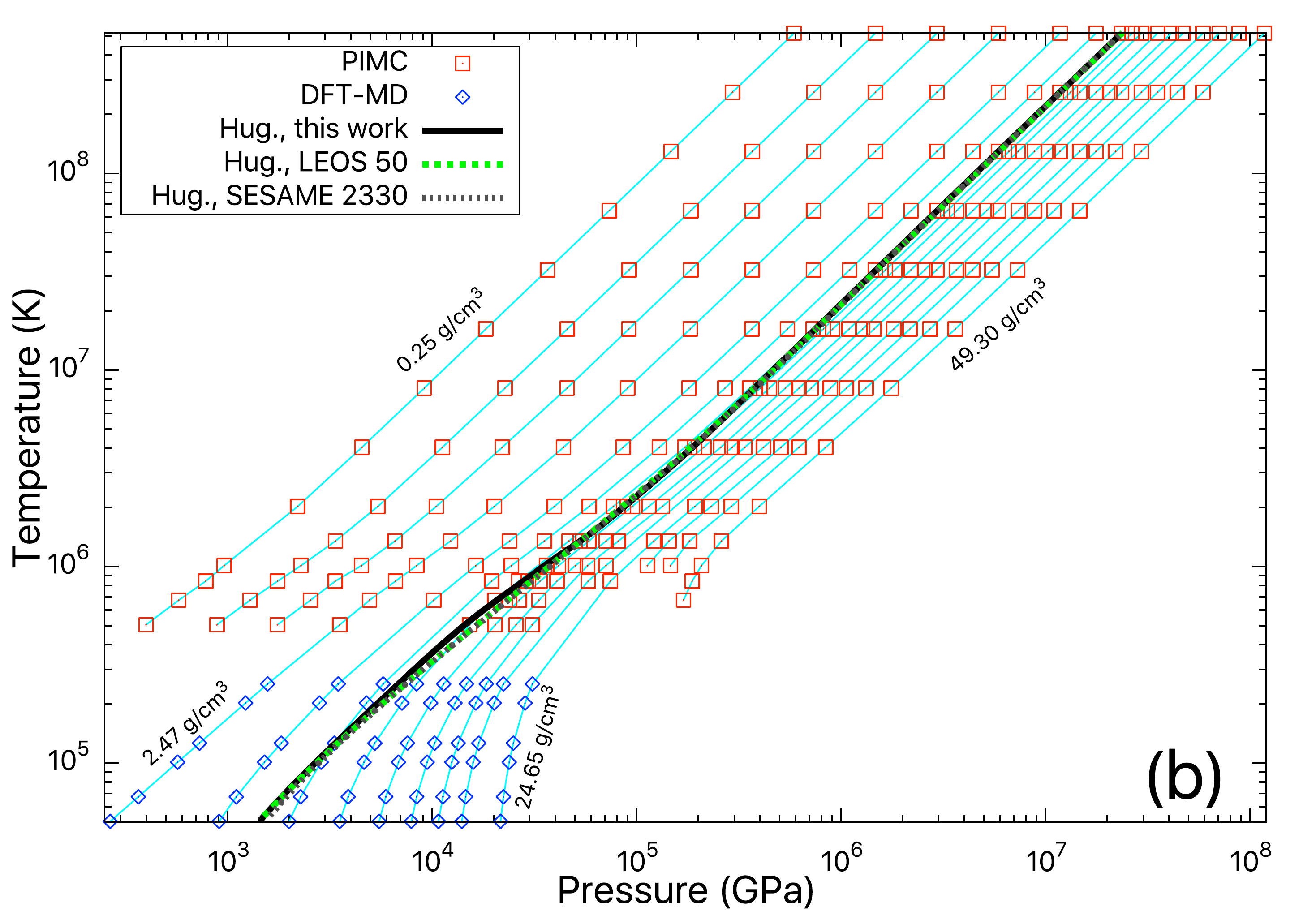}
\caption{\label{bhug} Boron EOS and shock Hugoniot curves shown in (a) $P-\rho/\rho_0$ and (b) $T-P$ plots. The Hugoniot curves from LEOS 50 and SESAME 2330 are co-plotted for comparison. Cyan-colored curves in panels (a) and (b) denote isotherms and isochores, respectively. The Hugoniot curves are constructed by choosing the initial density to be the same as $\rho_0$ ($\sim$2.46~g/cm$^3$).}
\end{figure}

We plot the Hugoniot curves thus obtained in a pressure-compression ratio 
($P-\rho/\rho_0$, where $\rho$ is the density in the shocked state) 
and a temperature-pressure ($T-P$) diagram
in Fig.~\ref{bhug}, and in a $T-\rho$ diagram in Fig.~\ref{Trhogrid}. 
Our EOS based on PIMC calculations predict a maximum compression
of 4.6 at 0.85 gigabar pressure and 2.0 million K temperature.
In comparison, LEOS 50 and SESAME 2330 models
predict boron to be stiffer by 6.9\% and 5.5\%, respectively, at the
maximum compression. The difference originates from the 1s shell
ionization, which increases the compression ratio and is well captured
in the PIMC simulations but not in the TF-based LEOS 50
and SESAME 2330 models. A similar deviation has been found for other
low-Z systems, such as CH~\cite{Zhang2017b,Zhang2018}.
At lower temperatures, LEOS 50 predictions
of the $P-\rho/\rho_0$ relation agree with our DFT-MD findings, while
SESAME 2330 predicts boron to be softer by 6-10\%. These are related
to the specific details in constructing the cold curve and the thermal ionic 
parts in the EOS models.

\begin{figure}
\centering\includegraphics[width=0.5\textwidth]{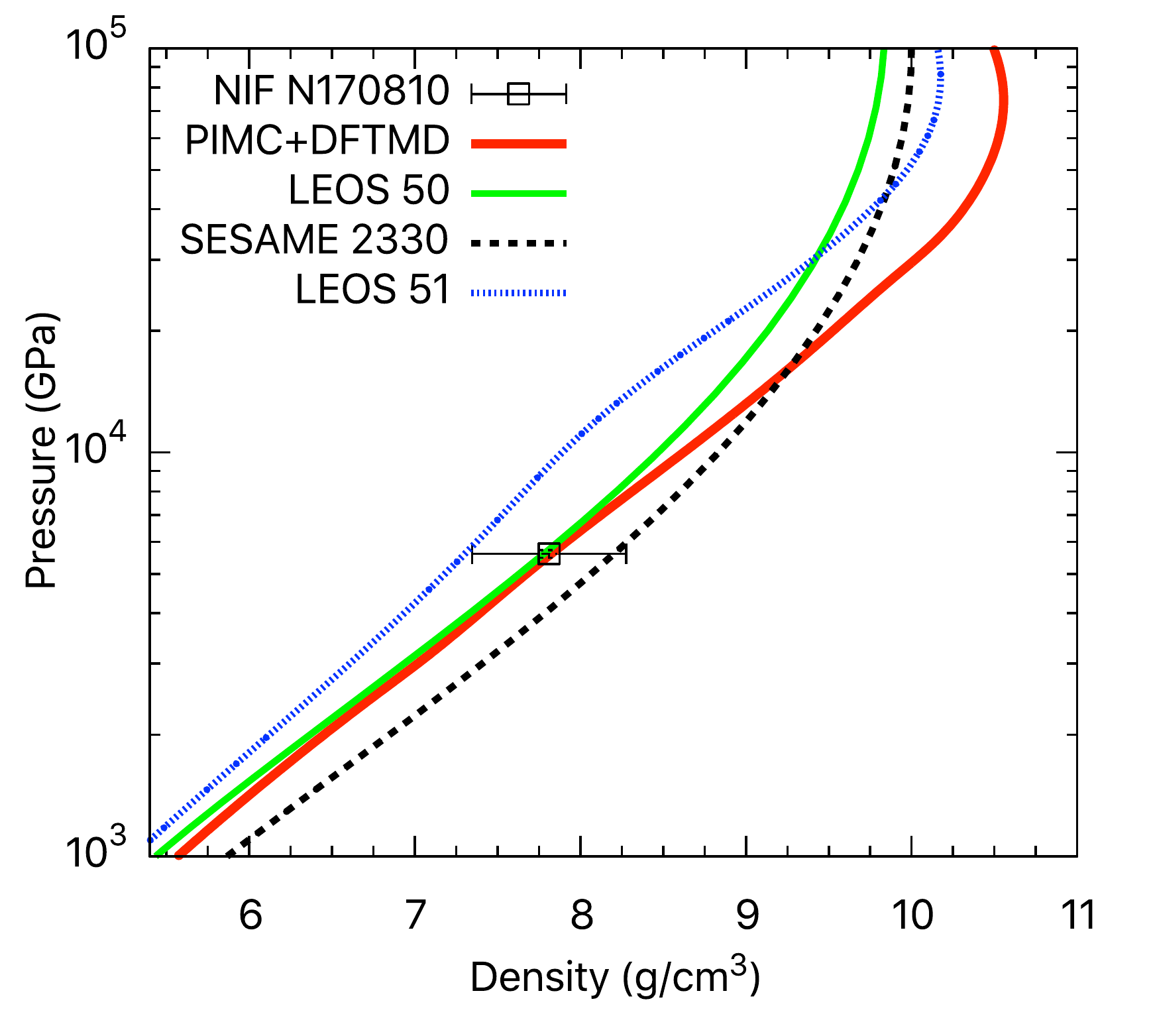}
\caption{\label{bhugPrho} Comparison of the experimental boron shock Hugoniot result with predictions from our first-principles EOS data and LEOS 50, SESAME 2330, and Purgatorio-based LEOS 51 models. When constructing the Hugoniot curve using the theoretical EOS data, the initial density is set to be the same as the experimental value of 2.31~g/cm$^3$ ($\beta$-boron).}
\end{figure}

The experimental boron Hugoniot data are summarized in Table~\ref{table:expdata}
and compared with our theoretical predictions in a pressure-density plot (Fig.~\ref{bhugPrho}).
The measured data point agrees perfectly with predictions by our first-principles calculations and LEOS 50,
but the predictions from the Purgatorio-based LEOS 51 and SESAME 2330 models are also consistent with
the measurement if the 1~$\sigma$ error bar in density is taken into account.

\section{Discussion}\label{discuss}
\subsection{Static and dynamic properties of boron plasmas}\label{struccharac}

\begin{figure}
\centering\includegraphics[width=0.5\textwidth]{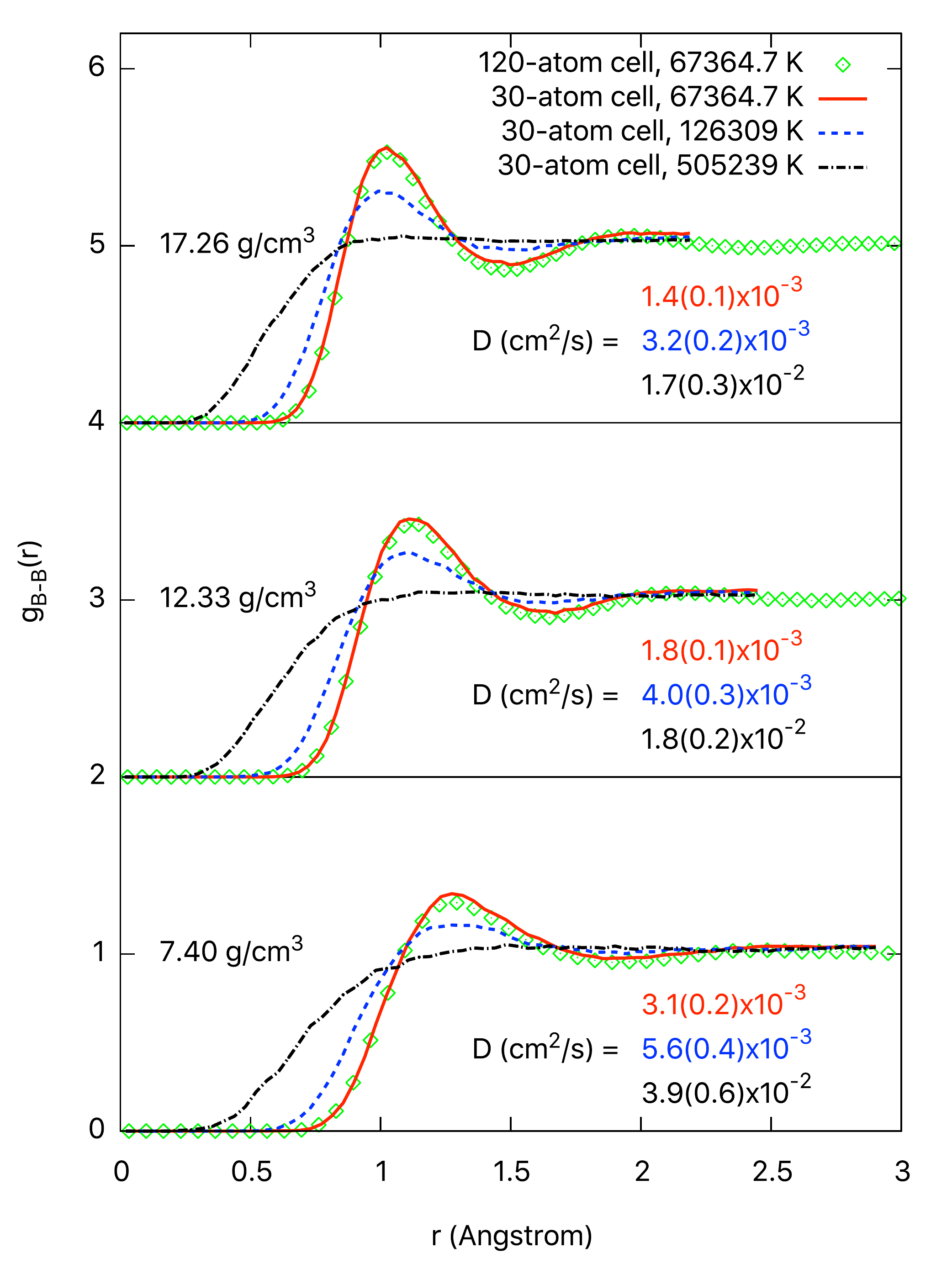}
\caption{\label{bgr} Nuclear pair correlation function $g(r)$ of boron at three densities and three temperatures. The $g(r)$ curves analyzed based on MD simulations of 120-atom cells at 6736.47 K are co-plotted for comparison. Curves at different densities have been off set for clarity. The consistency between $g(r)$ of 30- and 120-atom cells show negligible finite size effect in describing the ionic structures. The numbers at the inset of each panel show the values of self diffusivity at the corresponding density. Numbers in parentheses denote the standard error of the corresponding data. Red, blue, and dark-colored texts correspond to temperatures of 6.7$\times10^4$, 1.3$\times10^5$, and 5.1$\times10^5$~K, respectively.}
\end{figure}

The EOS and shock compression of warm and hot dense matter 
can be understood from the atomic and electronic structures.
Figure~\ref{bgr} compares the ionic radial distribution function $g(r)$ for boron
at selected densities (3-, 5-, and 7-times $\rho_0$) and temperatures (6.74$\times10^4$, 
1.26$\times10^5$, and 5.05$\times10^5$ K) from our DFT-MD simulations.
At 6.74$\times10^4$ K, the $g(r)$ function shows a peak-valley feature
between distances of 1.0--2.0 \AA{ } from the nucleus,
which is characteristic of a bonding liquid.
This feature gradually vanishes as temperature increases,
indicating that the system increasingly approaches an ideal gas.
However, there is a striking difference of the warm dense matter
from the ideal gas in that the atoms within this matter are partially ionized.

\begin{figure}
\centering\includegraphics[width=0.5\textwidth]{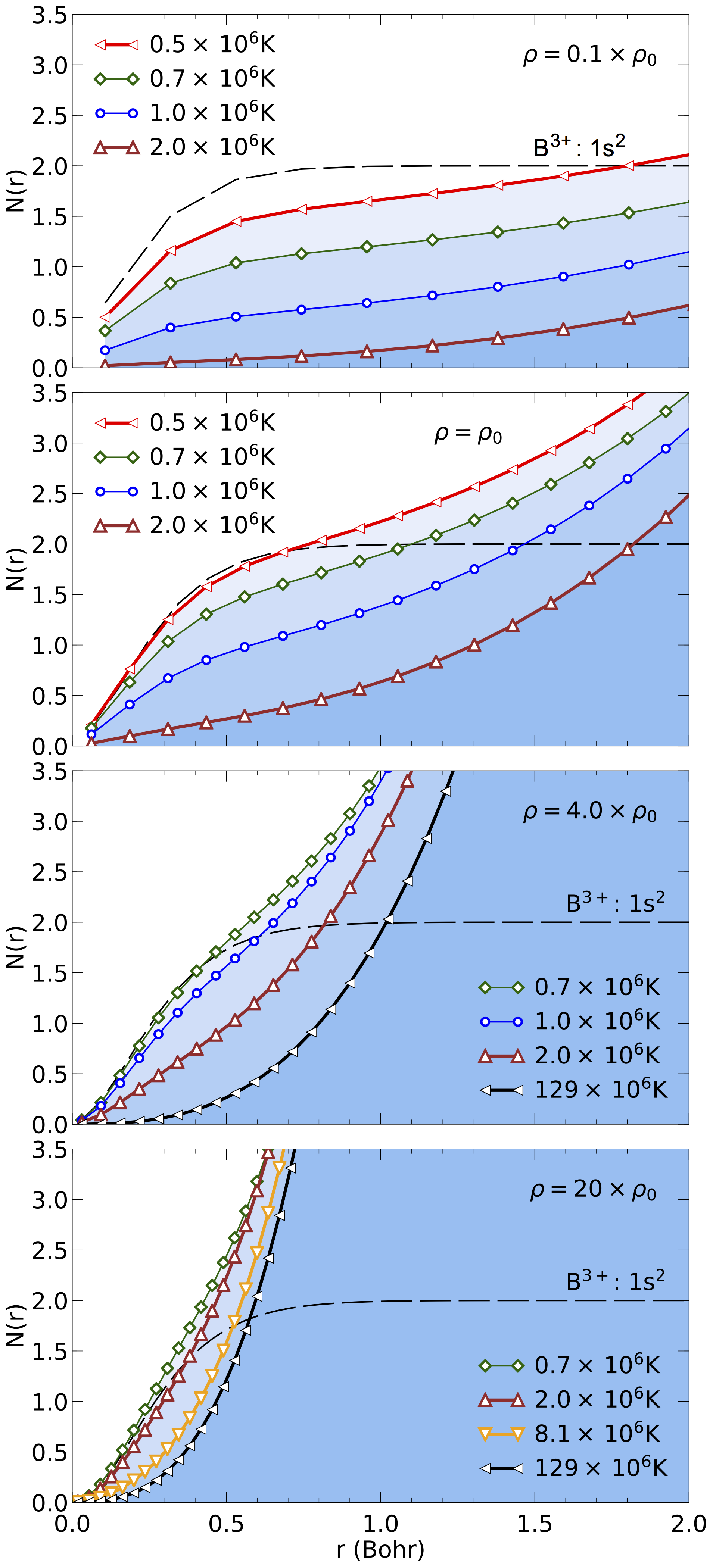}
\caption{\label{bNr} The average number of electrons around each nucleus at different densities and a series of temperatures. $\rho_0$ is 2.465~g/cm$^3$. The long dashed curve denotes the corresponding profile of the B$^{3+}$ ionization state calculated with GAMESS~\cite{gamess}. }
\end{figure}

The pressure-driven and thermal ionization processes can be well described by comparing
the $N(r)$ functions, which denote the average number of electrons within distance $r$ from
each nucleus, with the corresponding profile of the B$^{3+}$ ionization state.
$N(r)$ curves that are fully above the profile for B$^{3+}$ are associated with 
fully occupied $K$ shells, while those falling below indicate $K$-shell ionization.
The results at 0.1$\times$, 1.0$\times$, 4.0$\times$, and 20$\times\rho_0$
from our PIMC calculations are shown in Fig.~\ref{bNr}. We find no observable
ionization of the 1s states for $T$$<$0.5$\times$10$^6$ K at $\rho$$>$$\rho_0$, which
validates the use of the pseudopotential with a helium core in our DFT-MD simulations
in these temperature and density conditions. As $T$ exceeds $0.5\times10^6$ K, 
1s electrons are excited and thus contribute to the total pressure and energy
of the system,
which explains why both quantities are underestimated in DFT-MD,
as has been shown in Fig.~\ref{beosept} and discussed in Sec.~\ref{seceos}.

The $N(r)$ results also show that it requires higher temperatures for the $K$ shell
to reach the same degree of ionization at higher densities 
and that the same temperature change is associated with larger degrees of $K$ 
shell ionization at lower densities.  Previous 
generalized chemical models~\cite{Apfelbaum2013} showed increasing 
fraction of B$^{2+}$ particles at $T>3.5\times10^4$ K and negligible $K$ shell 
ionization within the complete temperature range (up to $4.2\times10^4$ K) of 
their study for low-density (0.094~g/cm$^3$) boron plasmas,
which remarkably agree with our findings here based on first-principles calculations.

\begin{figure}
\centering\includegraphics[width=0.5\textwidth]{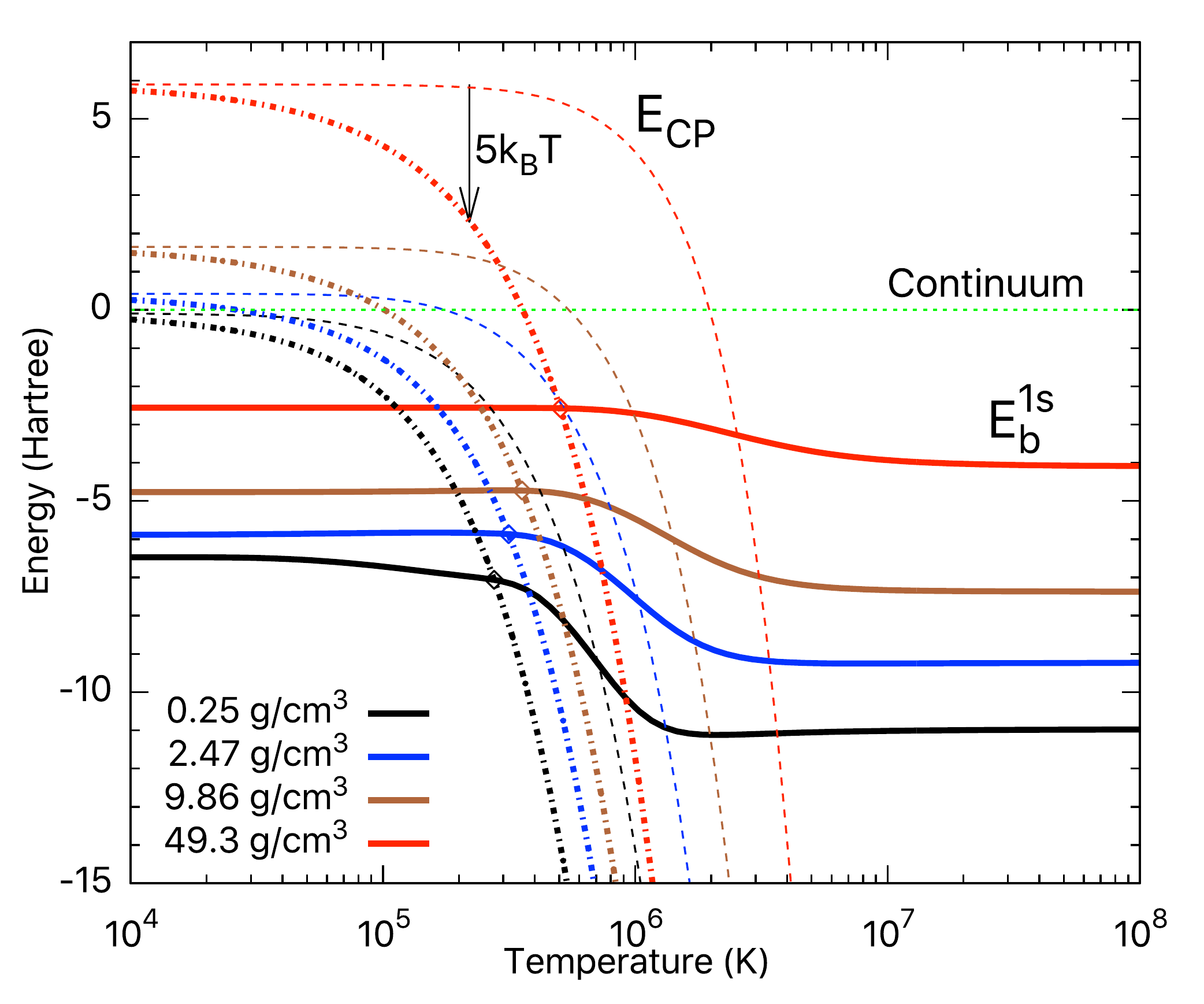}
\caption{\label{bEionization} Comparison of the 1s binding energy $E_\text{b}^\text{1s}$ (solid curves) with
the chemical potential $E_\text{CP}$ (thin dashed curves) as functions of temperature at four densities.
The data are obtained using the Purgatorio method. 
The dash-dotted curves represent $E_\text{CP}-5k_\text{B}T$.
The diamonds indicate the points at which the 1s level starts to be ionized (by 0.67\%).}
\end{figure}

In order to elucidate the physical origin of these observations, we compare the
temperature dependence of the 1s binding energy $E_\text{b}^\text{1s}$ with 
the chemical potential $E_\text{CP}$ along four different isochores between
0.1$\times$ and 20$\times\rho_0$.
The results are obtained using the Purgatorio method~\cite{Purgatorio2006,Sterne2007}
and are summarized in Fig.~\ref{bEionization}.
As density increases, $E_\text{b}^\text{1s}$ rises closer to the continuum level ($E$=0). 
$E_\text{CP}$ also increases with increasing density,
and in fact increases faster than $E_\text{b}^\text{1s}$. 
As a result, the Fermi occupation number of the 1s state 
actually increases with increasing density. 
At the temperature at which the $E_\text{b}^\text{1s}$ and $E_\text{CP}$ curves
intersect, the 1s energy level has a Fermi occupation number of 1/2.
The dash-dotted curves in Fig.~\ref{bEionization} plot the chemical potential 
minus $5k_\text{B}T$. The 1s level will have a Fermi occupation number of just 0.67\% 
below full occupancy at the temperature at which these curves intersect the 
corresponding 1s energy levels. This intersection therefore indicates the 
critical temperature at which the 1s level starts to ionize.
This intersection point shifts to higher temperature with increasing density, 
indicating that the ionization temperature increases with density, 
even though the 1s binding energy itself decreases. 
This accounts for the higher temperatures that are 
required for the $K$ shell to reach the same degree of ionization at higher densities,
as observed in Fig.~\ref{bNr}.
Purgatorio calculations of the $K$-shell occupation refines the critical temperature
to 3.2$\times10^5$--3.6$\times10^5$ K at densities between $\rho_0$--$4\rho_0$.
We have also compared the Purgatorio results to that of DFT simulations of 
boron on a face-centered cubic lattice using a dual-projector
Optimized Norm-Conserving Vanderbilt (ONCV)~\cite{oncv13,oncv13e}
pseudopotential with core radius equaling 0.8 Bohr. 
The ONCV and the Purgatorio results on chemical potential, $K$ shell ionization energies, 
and $K$ shell occupation are in good agreement with each other.

\begin{figure}
\centering\includegraphics[width=0.5\textwidth]{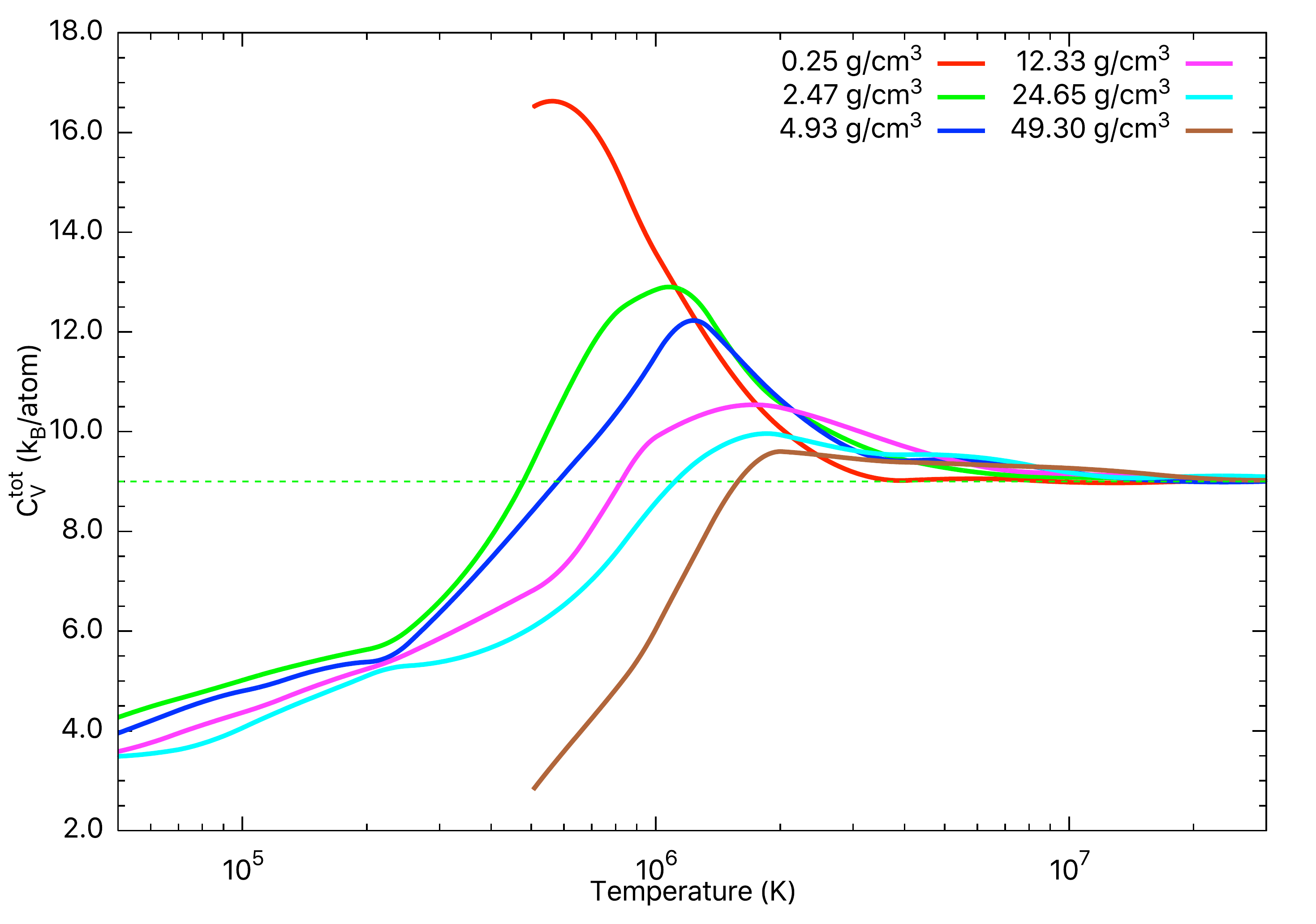}
\caption{\label{bCv} Total heat capacity $C_V^{\text{tot}}=\left(\partial E/\partial T\right)|_V$ of boron 
obtained from our DFT-MD and PIMC data along several isochores. 
All curves converge to the ideal-gas limit of 9$k_{\text{B}}$/atom at high temperature.
}
\end{figure}

The above findings about ionization are also consistent with the upshifting 
in energy, decreasing in magnitude, and expanding in width of the peak in 
heat capacity (Fig.~\ref{bCv}) as density increases. The peaks originate
from the excitation of 1s electrons of boron and appear at lower temperatures than
that of carbon in CH with comparable densities~\cite{Zhang2018}. This is 
because the $K$ shell of boron is shallower than that of carbon.

We also estimate the self diffusion coefficient $D$ for boron
using the mean square displacement and the Einstein relation.
We obtained values of $D$ that range between
$8\times10^{-4}$ and 0.05 cm$^2$/s at the temperatures 
($5\times10^4$--$5\times10^5$ K) and 
densities ($\rho_0$--$10\rho_0$)
that we performed DFT-MD simulations.
We find the values of $D$ (some shown in Fig.~\ref{bgr}) 
monotonically increase with temperature and the specific volume.
This is similar to what have been found for the diffusion of hydrogen 
in asymmetric binary ionic mixtures~\cite{Whitley2015} 
and deuterium-tritium mixtures~\cite{Kress2010}.

We note that accurate DFT-MD simulations of transport properties, 
such as diffusivity and viscosity, 
of one component plasmas across a wide coupling regime
are useful because of the potential breakdown of laws for ordinary 
condensed matter (e.g., the Arrhenius relation)~\cite{Daligault2006}. 
These studies also build the base for estimating the corresponding
properties of mixtures~\cite{Kress2010} which,
together with EOS approximations (e.g., average-atom or linear mixing 
approximation~\cite{Zhang2017b,Zhang2018}), are important in
characterizing multi-component plasmas.
However, such simulations require much more extended length 
of the MD trajectories and range of temperatures and densities in the 
more strongly-coupled regime, which are beyond the scope of
this work.

\subsection{PDXP performance sensitivity to EOS}\label{nifsens}

\begin{table*}
\centering
\caption{\label{PDXPtable} Polar direct-drive exploding pushers performance sensitivity to pressure change in boron EOS (based on LEOS 50). Corresponding data based on a GDP model are also shown for comparison.}
\begin{ruledtabular}
\begin{tabular}{c|c|c|c|c|c|c}
Pressure & Neutron & Xray& Gas Areal & Shell Areal & Convergence & Burn-averaged \\
Multiplier	& Yield 	&Bang Time (ns)	& Density (mg/cm$^2$)	& Density (mg/cm$^2$)	&Ratio & Ion Temperature (keV)\\
\hline \hline
0.8&	1.72$\times 10^{13}$&	2.22&	5.98&	3.11&	4.79&	20.48\\
1&	2.68$\times 10^{13}$&	2.24&	7.54&	3.82&	5.61&	22.11\\
1.2&	3.96$\times 10^{13}$&	2.28&	10.5&	5.06&	6.96&	21.73\\
\hline
GDP model from Ref.~\onlinecite{Ellison_2018}	&1.97$\times 10^{13}$	&3.02&	17.7&	23.8&	12.39&	7.11\\
\end{tabular}
\end{ruledtabular}
\end{table*}

In Ref.~\onlinecite{Ellison_2018}, a 1D ARES~\cite{arespaper1,arespaper2} model for the PDXP platform with GDP capsules was developed to match the x-ray bang time and yield of N160920-003, N160920-005, and N160921-001.  While 
we anticipate that changing the ablator in these experiments would necessitate recalibration of this model to match the performance of a new material, this model nonetheless offers a reasonable starting point for examining EOS 
sensitivity.  The capsule in N160920-005 consisted of a 18~$\mu$m thick GDP shell with an outer diameter of 2.95 mm, filled with 8-bar of D$_2$ gas and a trace amount of argon as a spectroscopic tracer.  The implosion was driven by a 
1.8 ns 
square pulse corresponding to a peak intensity of about $9.7\times 10^{14}$~W/cm$^2$.
The model developed in Ref.~\onlinecite{Ellison_2018} incorporates a multiplier on the energy delivered to the capsule, a flux limiter on the electron thermal conduction to account for inadequacies in the assumption of the diffusion model for 
heat transport, and a multiplier on the mass diffusion coefficient that is used to calibrate the multi-component Navier-Stokes model for mixing of the capsule ablator into the deuterium fuel.  The authors also modify the laser 
intensity used in the 1D simulations to account for geometric losses based on 2D ARES simulations.  
As 
discussed in Sec.~\ref{seceos}, our {\it ab initio} simulations yield pressures that differ by up to 20\% from the existing LEOS 50 table.  The largest variations occur at temperatures between about 
1$\times 10^5$ and 5$\times 10^6$K, as shown in 
Fig.~\ref{beosept}.  In this the regime, the electron thermal pressure is the largest contribution to the total pressure.  
We have therefore performed 1D ARES simulations using the LEOS 50 table with pressure multipliers of 0.8, 1.0, and 1.2 as a 
means of estimating the EOS sensitivity in a PDXP capsule using a boron ablator.  

Because boron is substantially more dense than GDP (2.465 g/cm$^3$ compared to 1.046 g/cm$^3$), and because the higher tensile strength should allow for a thinner shell, we have chosen a thickness of 6~$\mu$m for the boron capsules.  The results of 
the EOS sensitivity study are shown in Table~\ref{PDXPtable}.  We find that the variations in pressure considered here result in yield variations of -35\% to +48\%.  Higher ablator pressures result in higher gas areal density and higher 
convergence at burn time for very similar ion temperatures, thus the impact on yield is generated primarily via higher compression of the D$_2$ gas as the pressure in the ablator increases.  The shell areal density at the time of peak 
neutron production is also impacted by the pressure multiplier.  

For reference, the results from the model calculations in Ref.~\onlinecite{Ellison_2018} are also listed in Table~\ref{PDXPtable}.  We find that the 1D ARES model 
predicts lower gas and much lower shell areal density at peak burn time for the boron ablator compared to GDP.  This is because a larger portion of the thinner boron shell is ablated, allowing behavior more like a true exploding pusher than the 
thicker GDP ablator.  The GDP design has a substantial amount of unablated plastic, leading to a lower implosion velocity, higher convergence, and lower ion temperatures relative to the boron ablator.  
The first-principles calculations and experiments performed in this study will be used to generate a new EOS for B, which will be applied in future 2D calculations of the PDXP platform with a boron ablator.

\section{Conclusions}\label{conclusion}
In this work, we present first-principles EOS results of 
 boron using PIMC and DFT-MD simulations
from temperatures of 
5$\times10^4$ K to 5.2$\times10^8$ K. 
PIMC and DFT-MD cross-validates each other by showing 
remarkable consistency in the EOS ($<$1.5 Ha/B in total internal energy 
and $<$5\% in total pressure) at $5\times10^5$ K.
Our benchmark-quality EOS for boron provides an important 
base for future theoretical investigations of plasmas with boron.

We measured the boron Hugoniot at the highest pressure to date (56.1$\pm$1.2 Mbar)
in a dynamic compression experiment at NIF.
The result shows excellent agreement with that obtained from
the first-principles EOS data.
In addition, our calculations predict a maximum 
compression of 4.6, which originates from $K$ shell ionization and
 is slightly larger than those predicted by TF models 
LEOS 50 and SESAME 2330.

We investigated the PDXP performance sensitivity to
the EOS with a 1D hydrodynamic model.
The simulation results show that variations in pressure by -20\%
and 20\% result in neutron yield variations of -35\% to +48\%, respectively.

%\section{Supplementary Material}
%See the supplementary material for the EOS data table of boron from this study.

%----------------------------------------------------------------------
\begin{acknowledgments}
  
  This research is supported by the U. S. Department of Energy, grant
  DE-SC0016248. 
Computational support was mainly provided by the
  Blue Waters sustained-petascale computing project 
  (NSF ACI 1640776), which 
  is supported by the National Science Foundation 
  (awards OCI-0725070 and ACI-1238993) and the state of Illinois. 
  Blue Waters is a joint effort of the University of Illinois at 
  Urbana-Champaign and its National Center for Supercomputing Applications. 
S.Z. is partially supported by the PLS-Postdoctoral Grant of the 
Lawrence Livermore National Laboratory.
This work was in part performed under the auspices of the U.S. Department of Energy by Lawrence Livermore National Laboratory under Contract No. DE-AC52-07NA27344.
  
\end{acknowledgments}

%----------------------------------------------------------------------

%\bibliography{Boron.bib}
%\bibliography{Boron.bib}

%merlin.mbs apsrev4-1.bst 2010-07-25 4.21a (PWD, AO, DPC) hacked
%Control: key (0)
%Control: author (8) initials jnrlst
%Control: editor formatted (1) identically to author
%Control: production of article title (-1) disabled
%Control: page (0) single
%Control: year (1) truncated
%Control: production of eprint (0) enabled
%

\end{document}